\documentclass{article}

\usepackage{amsmath}
\usepackage{microtype}
\usepackage{graphicx}
\usepackage{subfigure}
\usepackage{booktabs} 

\usepackage{hyperref}

\usepackage[accepted]{mlsys2025}

\usepackage{multirow}
\usepackage{balance}

\newlength\savedwidth
\newcommand\whline{\noalign{\global\savedwidth\arrayrulewidth
                            \global\arrayrulewidth 1.5pt}%
           \hline
           \noalign{\global\arrayrulewidth\savedwidth}}

\newcommand{\passq}{{\tt pass-Q} }
\newcommand{\passkv}{{\tt pass-KV} }

\newcommand{\xxftodo}[1]{}

\renewcommand{\cite}[1]{\citep{#1}}

\begin{document}

\twocolumn[
\mlsystitle{Context Parallelism for Scalable Million-Token Inference}




\begin{mlsysauthorlist}
    \mlsysauthor{Amy (Jie) Yang}{to}
\mlsysauthor{Jingyi Yang}{to}
\mlsysauthor{Aya Ibrahim}{to}
\mlsysauthor{Xinfeng Xie}{to}
\mlsysauthor{Bangsheng Tang}{to}
\mlsysauthor{Grigory Sizov}{to}
\mlsysauthor{Jeremy Reizenstein}{to}
\mlsysauthor{Jongsoo Park}{to}
\mlsysauthor{Jianyu Huang}{to}
\end{mlsysauthorlist}

\mlsysaffiliation{to}{Meta Platforms, Inc., Menlo Park, California, USA}
\mlsyscorrespondingauthor{Jianyu Huang}{jianyuhuang@meta.com}

\mlsyskeywords{LLM, MLSys}

\vskip 0.3in

\begin{abstract}
        We present context parallelism for long-context large language model inference, which achieves near-linear scaling for long-context prefill latency with up to 128 H100 GPUs across 16 nodes. Particularly, our method achieves 1M context prefill with Llama3 405B model in 77s (93\% parallelization efficiency, 63\% FLOPS utilization) and 128K context prefill in 3.8s. We develop two lossless exact ring attention variants: {\tt pass-KV} and {\tt pass-Q} to cover a wide range of use cases with the state-of-the-art performance: full prefill, persistent KV prefill and decode. Benchmarks on H100 GPU hosts inter-connected with RDMA and TCP both show similar scalability for long-context prefill, demonstrating that our method scales well using common commercial data center with medium-to-low inter-host bandwidth.

\end{abstract}
]




\printAffiliationsAndNotice{}  

\section{Introduction}
\label{introduction}

Contemporary large language models (LLMs), such as Llama~\cite{llama1,llama2,llama3}, Gemini~\cite{team2023gemini,reid2024gemini}, GPT-4~\cite{achiam2023gpt}, require significant computational resources for inference, especially with long context lengths:
OpenAI GPT-4o 128K context length~\cite{openai_gpt4}, Anthropic's Claude with 200K context length~\cite{anthropic_claude}, Google's Gemini 1.5 Pro with 1M context length~\cite{google_gemini}.
With a single H100 GPU host (8 GPUs), it can take 60 seconds to serve 128K context length\footnote{Google's Gemini 1.5 Pro (Sep 2024) has a latency of 20.43 seconds for time to first token on 100K context length, from \url{https://artificialanalysis.ai/models/gemini-1-5-pro/prompt-options/single/100k}.} or 1200 seconds to serve 1M context length for Llama3 405B model.
Context parallelism (CP) is a system optimization technique that improves the latency and scalability of LLM inference, particularly for long contexts. Without modifying the underlying dense attention algorithms, CP offers several advantages for long-context LLM inference:
\begin{itemize}
    \item \textbf{Compute parallelization}: CP distributes computation across multiple GPUs in order to reduce latency, in contrast with pipeline parallelization (PP)~\cite{huang2019gpipe} that improves throughput but not latency.
    \item \textbf{Communication message size reduction}:
        Compared to tensor parallelism (TP)~\cite{shoeybi2019megatron}, CP demands less communication bandwidth in multi-host environments, by maintaining a communication size that is orders of magnitude smaller than TP, especially for inter-node communication.
    \item \textbf{KV cache distribution}: Key and value (KV) embeddings grow linearly with context length. CP distributes the storage of KV embeddings across multiple GPUs, enabling larger batch sizes with the addition of more CP ranks.
\end{itemize}

To the best of our knowledge, this is the first paper to disclose the system implementation details on applying context parallelism in inference scenario.
Our main contribution lies in the adaptation and optimization of ring attention~\cite{liu2023ring} for efficient LLM inference with long context lengths. While the previous work primarily focuses on leveraging ring attention to enhance training throughput for long sequences, this paper identifies and addresses unique challenges posed by inference:


\begin{itemize}

    \item \textbf {Support for multi-turn prefill and decoding}: We recognize the importance of multi-turn conversations, a common characteristic of online LLM applications.
        Unlike prior research focused on training, we introduce novel strategies on load-balanced sharding for persistent KV cache and parallelization algorithms that leverage sharded KV cache across multi-turn prefill and decode. These mechanisms are crucial for maintaining conversation history during inference.

    \item \textbf {Optimization for latency}: Latency is critical for user experience in real-time inference. To optimize latency in multi-turn conversations, we developed {\tt pass-KV} and {\tt pass-Q} ring attention variants and heuristics to dynamically select the ring attention algorithms for the lowest latency under varying context lengths and KV cache hit rates.
    \item \textbf {Compute and memory load balancing}: To maintain balanced load among CP ranks across batched requests with varying input lengths, we introduce load-balanced sharding of both input tokens and KV cache entries. Previous work targets training typically with uniform sequence length. We proposed innovative algorithms to ensure even distribution of compute and KV cache memory across CP ranks, contributing to improved overall performance and scalability.

\end{itemize}

In essence, our work extends context parallelism to efficiently address the challenges and requirements of serving millions of tokens in LLM inference. We introduce novel algorithms and heuristics for optimizing ring attention, demonstrating their effectiveness in reducing latency, improving KV cache utilization, and enabling scalable distributed inference for long-context LLMs.
Since our method focuses on system-level optimizations, it can be seamlessly integrated with architectural innovations or algorithmic enhancements to further amplify performance gains.





\section{Background}

\subsection{Large Language Models (LLM)}
Since the introduction in the seminal work \cite{vaswani2017attention}, the transformer model architecture has become the fundamental building block for modern language models. Recently, language models have increased exponentially in complexity (measured in number of parameters). Examples: BERT was trained with 0.34B parameters in 2018 \cite{devlin2018bert}, 1.5B parameter GPT-2 was released in 2019 \cite{radford2019language}, and 175B parameter GPT-3 was released one year later in 2020 \cite{brown2020language}, and the latest Llama 3.1 model pushed to 405B parameters \cite{llama3}.

Besides the parameter number, the \emph{context length} is another important indicator of LLM's capabilities. In general, a longer context window indicates better capability to handle a large body of input texts, audios, images, and videos. Modern LLMs support 128K to more than 1M  context lengths \cite{openai_gpt4, anthropic_claude, google_gemini}.

\subsection{Challenges with Serving Long Context LLM}
In this work, we mainly address the challenges with extremely large (128K-1M) context lengths.
\begin{itemize}
    \item \textbf{Compute}: 
    While an $ W$-parameter Transformer model requires $ 2 \cdot W $ matrix multiplication FLOPs for each token during inference or forward pass \cite{kaplan2020scaling}, the pairwise attention architecture found in mainstream transformers \cite{vaswani2017attention} incurs a quadratic cost in FLOPs w.r.t. context lengths, which would be dominating in long context cases. Several approximate and sparse methods were proposed, including focusing attention on a subset of tokens, and employing a combination of local and global attention strategies. Techniques such as window attention~\cite{liu2021swin}, local attention~\cite{xiong2021simple}, Linformer~\cite{wang2020linformer}, and semi-local sparse attention~\cite{jiang2024minference,beltagy2020longformer} are examples of such innovations that help manage the computational cost.

    \item \textbf{Memory}: 
Memory usage for LLMs, particularly the KV cache~\cite{pope2023efficiently}, scales linearly with the context length. Model compression techniques such as KV cache quantization are crucial for bending the growth curve: lower precision formats like 3-bit, INT4/8 or FP8 can achieve a $ 2 \times $ to $ 4 \times $  reduction in memory requirements compared to using 16-bit~\cite{hooper2024kvquant,lin2024qserve}. Grouped Query Attention (GQA)~\cite{ainslie2023gqa} and MQA~\cite{shazeer2019fast} were widely adopted to reduce memory usage by reducing the number of KV heads by $ 8 \times $ to $ 64 \times $. Additionally, strategies like paged attention~\cite{kwon2023efficient} have been developed to provide efficient page-like memory management for large numbers of tokens.

\end{itemize}

\subsection{Prior works on Long Context LLM}
The following are the main directions to achieve efficient long context window LLM inference:
\begin{itemize}
    \item \textbf{New model architectures}: introduce long context window comprehension components at pretraining stage~\cite{munkhdalai2024leave}.
    \item \textbf{Post-training changes}: modify a pretrained model with shorter context window to support longer or even infinite context windows \cite{xiao2023efficient}.
    \item \textbf{System-level optimizations}: preserve the model architecture, instead improve the scalability of existing dense attention algorithms to leverage more compute resources \cite{li2021sequence,brandon2023striped,liu2023ring,wu2024loongserve,li2023distflashattn,jacobs2023deepspeed,fang2024usp}.
\end{itemize}

Our work falls into the third category, and can be used in conjunction with methods from the other two categories with minor or no modifications. Our method accelerates future algorithmic research or real-world LLM applications for long-context LLM serving, and also provides the flexibility to trade off model inference latency with hardware capacity depending on the latency requirements of specific applications.


\section{Context Parallel Inference}

\subsection{Notations}
\label{sec:notation}
The notations used in this paper are summarized in Table \ref{tab:notation}.

\begin{table}[t]
\caption{Notation table.}
\vskip 0.05in
\centering
{\footnotesize
\setlength{\tabcolsep}{3pt}
  \begin{tabular}{c | l }
  \whline
  Notation & Description  \\
  \whline
      $ N_H $ &  Number of query heads (or attention heads) \\
\hline
$ N_{KV} $ &  Number of key/value heads \\
\hline
$ D_H $ & Head dimension in Transformer \\
\hline
$ D $ & Model dimension in Transformer: $ D_H \cdot N_H $ \\
\hline
$ Q, K, V $ & Query, Key, Value tensors \\
\hline
$ B $ & Batch size (\# of input sequences)  \\
\hline
$ W $ & Model parameter size \\
\hline
$ T $ & Input sequence length \\
\hline
$ P $ & Previously cached KV length \\
\hline
$ e $ & Element data type size \\
\hline
$ C $ & Peak compute \\
\hline
$ BW $ & Peak comm bandwidth  \\
\hline
$ bid $ & Decode batch id \\
\hline
      \multirow{2}{*}{ $ T^i_{j}  |  P^i_{j} $ } & \# of \{new tokens $ | $ cached KV tokens\} in $ i $-th  \\
                                                   & sequence sharded to CP rank $ j $ \\
\hline
      $ L^i $& $ KV $ length in $ i $-th sequence: {\tiny $ max_{j=0}^{N-1}(P^i_j+T^i_j) $} \\
\hline
$ N $ & Number of hosts/nodes/CP ranks \\
\hline
$ N_{TP} $ & TP group size (\# of GPUs in a TP group) \\
\hline
      \multirow{2}{*}{ $ Q^s_{k} | KV^s_{k} | bid^s_{k} $} & \{Query $ | $ key and value tensors $ | $ batch id \} \\
      & on rank $ k $, originally allocated to rank $ s $ \\
\hline
      $ O^s_{k} $& Attention output from $ Q_k $ and $ KV^{s} $ \\
\hline
      \multirow{2}{*}{$ TP8,TP16 $} & Tensor parallel sharding over 8 GPUs  \\
      & on one node or 16 GPUs on two nodes \\
\hline
      \multirow{2}{*}{ $ CP_N $} & Context parallel sharding on $ N $ nodes  \\
      & with TP8 for each node (same as {\tiny $ CP_N $+$ TP8 $}) \\
\hline
      \multirow{2}{*}{ $ TTFT $} & Time-to-first-token: latency for prefilling \\
      & the whole input tokens \\
\hline
      \multirow{2}{*}{ $ TTIT $} & Time-to-incremental-token: latency for \\
      & decoding each output token \\
  \whline
  \end{tabular}
}
\label{tab:notation}
\end{table}


\subsection{Model Parallelization}


Large language models are commonly parallelized across multiple GPUs using a combination of various parallelism paradigms:
\textbf{Tensor Parallelism (TP)}~\cite{shoeybi2019megatron,korthikanti2023reducing} partitions the weights of fully connected layers (i.e., linear layers) by alternating the sharding in row and column dimensions.
\textbf{Pipeline Parallelism (PP)}~\cite{narayanan2021efficient} shards layers into different pipeline stages, and splits input tensors along the batch size dimension into micro-batches to orchestrate a pipeline schedule to optimize the system throughput.
Instead of sharding model weights, \textbf{Context Parallelism (CP)}~\cite{li2021sequence} distributes input tokens to multiple GPUs along the sequence length dimension. CP ranks communicate QKV tensors for attention, which is the only computation with dependency between tokens in the same sequence. 

Both TP and CP reduce latency when scaled to multiple nodes. Compared with TP, CP provides an alternative design choice for trade-offs between memory consumption and system performance.
As detailed in Table~\ref{tab:TP-CP-msg-size-table}, CP communicates token embeddings on attention layers while TP communicates on linear layers.
CP has less communication traffic for two reasons:
(1) Contemporary LLMs have more linear layers than attention layers: each canonical transformer block has four linear layers and one attention layer.
(2) CP may communicate KV tensors instead of Q tensors, which leads to much less communication for models with GQA \cite{ainslie2023gqa}. For Llama3 405B model with 128 query heads and 8 KV heads ($ N_{KV} = 8 $ vs. $ N_H = 128 $), communicating KV heads has $ 16 \times $ smaller message sizes than communicating query heads \cite{llama3}.
CP's communication cost advantage over TP results in significant latency improvements for multi-node inference,
as interconnect bandwidth between nodes are several times lower than intra-node bandwidth (Section \ref{sec:cp-vs-tp}).
Although CP offers lower communication costs, it incurs higher memory consumption because its lack of model weight sharding.

In this paper, we design and implement an efficient LLM inference system with CP to unblock such a trade-off when scaling out the number of GPUs.
In practice, we set TP size into a number (usually 8) to fit the model into GPU memory, and we leverage CP to efficiently scale out into multiple nodes as it saves communication traffic.

\begin{table}[t]
    \caption{Communication and memory cost comparison between tensor parallel (TP) and context parallel (CP) for full prefill. $ T $: sequence length, $ D_H $: head dimension, $ N_H $: \# of attention heads, $ N_{KV} $: \# of key/value heads, $ N_{TP} $: TP group size, W: model parameter size. Total comm cost shows the communication cost per transformer block.} 
\vskip 0.15in
{\footnotesize
\begin{center}
\begin{sc}
\setlength{\tabcolsep}{3pt}
\begin{tabular}{l | c | c}
  \whline
~ & TP & CP \\
  \hline
{Collective}          & AllReduce & SendRecv \\
{Comm per 2 Linear}  & $ T \cdot N_H \cdot D_H $ & 0 \\
{Comm per Attn} & 0 &  $ T \cdot N_{KV} \cdot D_H $ \\
{Total comm} & $ 2 \cdot (T \cdot N_H \cdot D_H) $ & $ T \cdot N_{KV} \cdot D_H $ \\
  \hline
{Parameter Size} & $\frac{W}{N_{TP}}$ & $W$ \\ [0.4em]
  \whline
\end{tabular}
\end{sc}
\end{center}
}
\label{tab:TP-CP-msg-size-table}
\vskip -0.1in
\end{table}

\subsection{Inference Prefill and Decode Attention}

We characterize large language model online inference for multi-turn messaging into three stages: full prefill, partial prefill, and decode. When user initiates the conversation with an initial prompt, the entire prompt goes through \textbf{full prefill}, where we compute full causal attention between tokens. Projected key and value tensors from multi-head (MHA) or grouped query attention (GQA)~\cite{ainslie2023gqa} are saved in GPU HBM as KV cache. After the initial full prefill, the model then starts generating a response with auto-regressive \textbf{decoding}, where a new token attends to previously cached KV tensors and outputs response tokens one at a time. KV values generated during decoding stage are also saved in KV cache. After the server returns a response, the user may give a follow-up prompt, which will go through \textbf{partial prefill} (or \textbf{persistent KV prefill}), where tokens within the new prompt attend to themselves as well as all cached tokens in the previous prompt and model response. This process may repeat multiple times in real world applications, which requires persistency of KV cache between prompts from the same user.

\subsection{Computation and Communication Modeling}
Each of the three stages of multi-turn online LLM inference carries different performance characteristics.

Assume we have an input sequence with length $T$, with previously cached KV length $P$, and a generic GQA model with $N_H$ query heads, $N_{KV}$ key and value heads and model dimension $D$. We have the following shapes for query (Q), key (K), and value (V) embeddings:
$$
shape(Q) = [T, N_H, \frac{D}{N_H}]
$$
$$
shape(K) = shape (V) = [(T + P), N_{KV}, \frac{D}{N_H}]
$$

When Q and KV have the same lengths, passing KV around in ring attention incurs smaller traffic than passing Q, and the communication can be fully overlapped with attention computation \cite{li2021sequence}. LLM training guarantees this property $ len(Q) = len(K) = len(V) = T $, or equivalently, $ P=0 $.
This is not necessarily true for inference as $ len(Q) $, $ len(K) $, and $ len(V) $ depend on user behaviors and KV cache configurations.

For inference, with high persistent KV hit rate, the ring attention algorithm that always passes KV around may not provide the best performance, as:
\begin{itemize}
    \item Attention computation is much faster with fewer Q than cached KV. Communication cost will be exposed on critical path if not fully overlap with computation.
    \item When Q is significantly smaller than the cached KV, communicating the full persistent KV would be significantly more costly than communicating Q.
\end{itemize}

To achieve better inference performance for full prefill, persistent KV prefill, and decode, we extend ring attention with an option to pass Q instead of KV, when passing Q leads to less communication cost. Specifically, Q embeddings have smaller size than KV embeddings if:

\begin{equation}
\label{eqn:pass-q-vs-kv}
\frac{T}{T+P} \leq 2 \cdot \frac{N_{KV}}{N_H}
\end{equation}

Note that the right hand side (RHS) is constant given a pretrained model. Therefore we can use the RHS as a constant threshold to switch between passing KV embeddings and Q embeddings dynamically depending on ${T\over T+P}$, or the \emph{KV cache miss rate} ( $ 1 - $ \emph{KV cache hit rate}).

Specifically, for full prefill where $ P = 0 $, communicating KV embeddings results in a smaller message size for GQA models with $ N_H > 2 \times N_{KV} $. For decoding where $ T=1 $, communicating Q embedding almost always results in smaller communication sizes. Consequently, we leverage ring \passkv{} for full prefill, and ring \passq{} for decode and partial prefill with high KV cache hit rate.

To understand whether communication can be reliably overlapped with attention computation with varying persistent KV hit rates, we approximate the attention computation and QKV communication latency using a simple roof-line model (Table \ref{tab:roofline}).

\begin{table}[t]
    \caption{GQA attention complexity for full prefill and partial prefill ($e$: number of bytes per element).}
\label{partial-prefill-gqa-complexity}
\vskip 0.15in
\begin{center}
\begin{small}
\begin{sc}
\begin{tabular}{l | c | c}
\whline
~ & full prefill & partial prefill \\
\hline
FLOPS          & $4 T^2 D$  &  $4 T D (T + P)$  \\
Q bytes        & $T D e$ & $T D e$  \\
    KV bytes       & $2 T D \frac{N_{KV}}{N_H} e$ & $2 (P +T) D \frac{N_{KV}}{N_H} e$  \\ [0.5em]
\whline
\end{tabular}
\end{sc}
\end{small}
\end{center}
\label{tab:roofline}
\vskip -0.1in
\end{table}

Let's assume a system with peak compute of $C$, bandwidth of $ BW $ for QKV communication, new token length $T$, and cached token length $P$. We focus the analysis on prefill with low persistent KV hit rate, which is compute-bound and the culprit of long (e.g. 60s) prefill latency for inference. In the following analysis, we aim to identify values of $P$ and $T$ such that the communication latency is smaller than the computation latency. In simplified terms: $\frac{FLOPS}{C} \geq \frac{min(Q_{bytes}, KV_{bytes})}{BW} $.

For low-to-medium KV cache hit rate prefill, we will not be bound by ring {\tt pass-KV} communication if:
$$
\frac{4\cdot T \cdot D (T + P)}{C} \geq \frac{2 \cdot (T + P) \cdot D \cdot e \cdot \frac{N_{KV}}{N_H}}{BW}
$$

To extend to multi-host distributed inference, we would further partition each CP rank with TP over intra-node GPUs, and add additional CP nodes to increase parallelization on context dimension. For CP over $ N $ nodes, we would be able to hide ring {\tt pass-KV} communication latency under attention computation if:
$$
\frac{4\cdot T \cdot D (T + P)}{N \cdot C} \geq \frac{2 \cdot (T + P) \cdot D \cdot e \cdot \frac{N_{KV}}{N_H}}{BW}
$$

\begin{equation}
\label{eqn:pass-kv-overlap}
    T \geq N \cdot \frac{C \cdot {N_{KV}} \cdot e}{2 \cdot {N_H} \cdot BW}
\end{equation}


Note that the threshold for $ T $, the length of new tokens is a static threshold with respect to a given model and hardware, which is independent of KV cache hit on the previously cached KV length $ P $.

Similarly, in a distributed inference setting with CP over $ N $ nodes,
we will not be bottlenecked by ring {\tt pass-Q} communication if:
$$
\frac{4 \cdot T \cdot D (T + P)}{N \cdot C} \geq \frac{T \cdot D \cdot e}{BW}
$$
\begin{equation}
\label{eqn:pass-q-overlap}
(T + P) \geq N \cdot \frac{e \cdot C}{4 \cdot BW}
\end{equation}

Note that RHS is also static with respect to one particular system. As we have discussed, we will leverage \passq{} when the number of new tokens to prefill $T$ is significantly smaller than the number of cached tokens $P$. In this case, whether we will be able to completely overlap the latency for communicating Q is determined by the total context length $ (T + P) $. Sufficiently large total context length would allow us to overlap the {\tt pass-Q} communication regardless of KV cache hit rate.



\begin{algorithm}[h]
    \caption{Pass-KV vs. Pass-Q Partial Prefill Heuristics}
   \label{alg:pass-kv-vs-pass-q-partial}
\begin{algorithmic}
   \IF{
        $T \geq N \frac{C \cdot {N_{KV}} \cdot e}{2 \cdot {N_H} \cdot BW} $ or $  \frac{T}{T+P} \geq 2 \frac{N_{KV}}{N_H} $
   }
   \STATE pass-KV
   \ELSE
   \STATE pass-Q
   \ENDIF
\end{algorithmic}
\end{algorithm}

To summarize, we adaptively switch between {\tt pass-KV} and {\tt pass-Q} for inference \textbf{partial prefill} following the heuristics in Algorithm \ref{alg:pass-kv-vs-pass-q-partial}\footnote{In practice, the achieved BW and C are lower than the theoretical hardware peaks. We start with these peak values and then fine-tune the thresholds based on empirical data.}.
It's worth noting that the \textbf{full prefill} can be considered as a special case where $ P = 0 $, while \textbf{decoding} can be viewed as a special case where $ T = 1 $.
We can calculate the static thresholds for this heuristics once based on the system and model spec, and use the heuristics to choose which options to use dynamically for the optimal performance in a wide combination of total context length and KV cache hit thresholds.

\subsection{Ring Pass-KV, Pass-Q Prefill}
We implemented both {\tt pass-KV} and {\tt pass-Q} ring attention to minimize the communication latency with different context lengths and KV cache hit rate. In this section, we delve into the implementation details for achieving effective load balancing and communication overhead management, which are critical to the the scalability of distributed context parallel inference.
    \subsubsection{Load Balanced Sharding} \label{section:load-balanced-sharding}
    In causal attention each token attends to all tokens before it in the same sequence. Naively partitioning all tokens evenly over CP ranks in the order of the original sequence results in imbalanced compute over different CP ranks. Prior work leverages order permutation and uneven partition to achieve load balance for causal attention \cite{cho2024kv, brandon2023striped}. To support maximum context length provided by the pretrained model without OOM on any particular CP rank with heavier load, we aim for load-balancing for both attention compute and KV cache capacity. To shard an input sequence into $N$ CP ranks, we partition the sequence evenly into $2 \times N$ chunks: $ C_0, C_1, ..., C_{2 \times N -1} $, and have each CP rank $i$ take two chunks: ($C_i$, $C_{2 \times N - i - 1}$).

\begin{figure}[t]
    \centering
    \includegraphics[width=8cm]{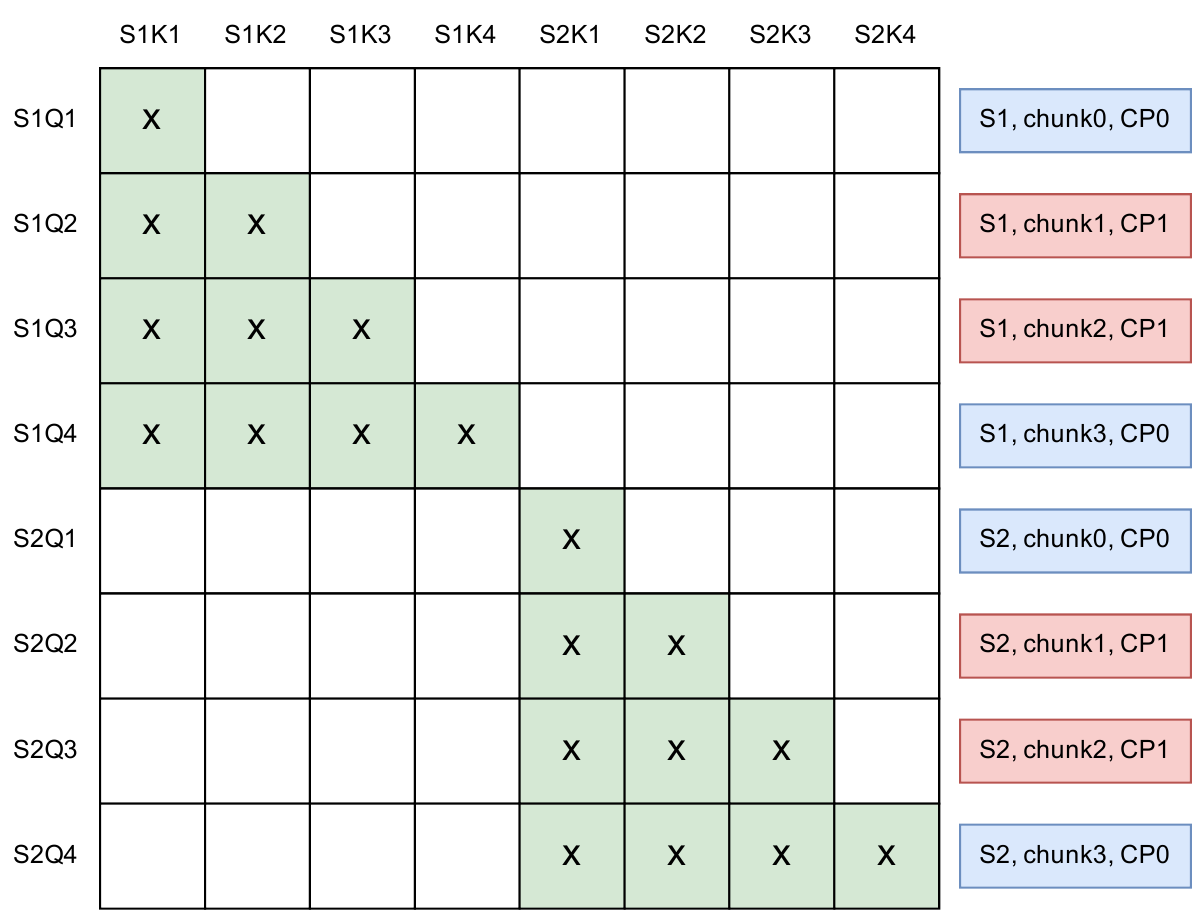}
        \caption{Load-balanced CP sharding with fused inputs in full prefill with 2 CP ranks (CP2). We have 2 input sequences: $ S1 $, $S2 $. Each is partitioned evenly into 4 chunks: $Q_i$ / $K_i$, where $i=1, 2, 3, 4$.}
    \label{fig:cp-sharding}
    \end{figure}

    For fused variable length inputs in full prefill, we partition each individual sequence in the same way and pad the input sequence length if needed (Figure \ref{fig:cp-sharding}).

\begin{figure}[t]
    \centering
    \includegraphics[width=8cm]{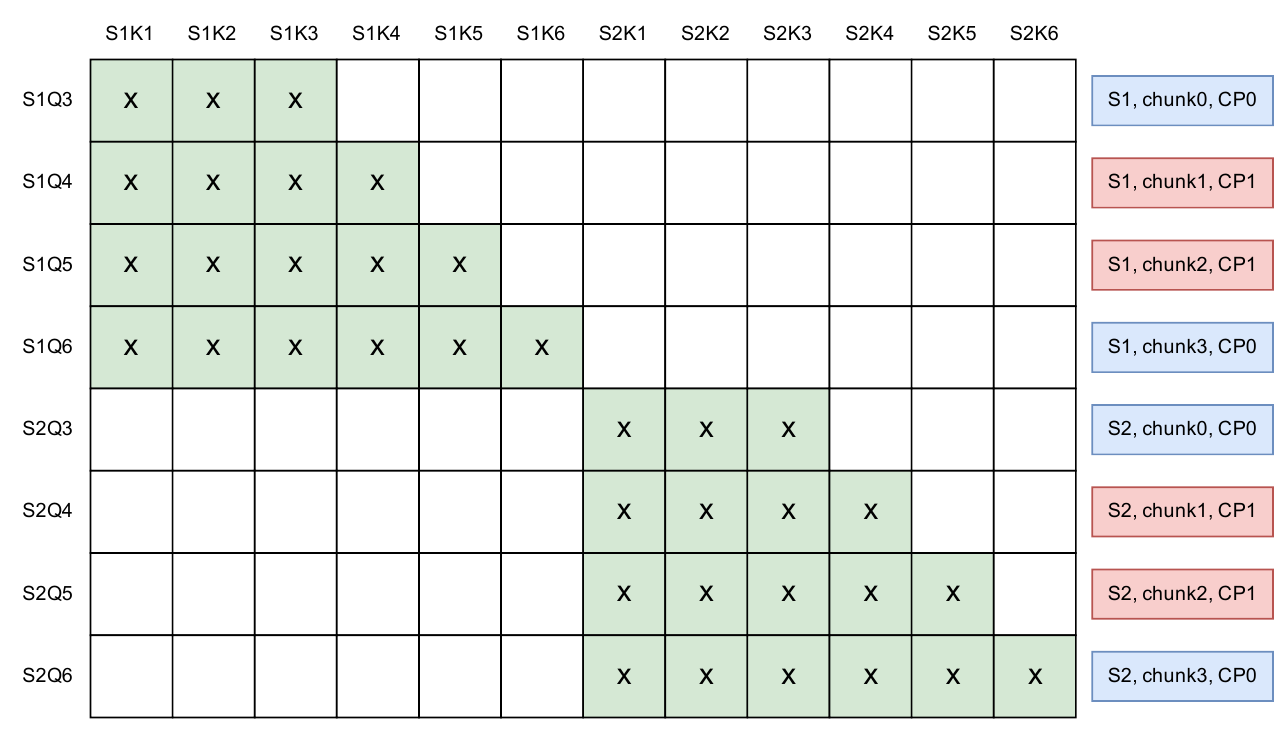}
        \caption{Load-balanced CP sharding with fused inputs partial prefill with 2 CP ranks (CP2). We have 2 input sequences: $ S1 $, $ S2 $. Load-balanced sharding is applied to the new token $ Q_i $ dimension (4 chunks), regardless of how cached token dimension $ K_i $ is partitioned in partial prefill.}
    \label{fig:cp-sharding-partial}
    \end{figure}

    For partial prefill with new tokens (total length: $ T $) and cached tokens (total length: $ P $), we apply the load-balanced sharding in the dimension of the new tokens regardless of cached tokens (Figure \ref{fig:cp-sharding-partial}).

    \subsubsection{Ring Pass-KV Algorithm}
    \label{sec:ring-pass-kv-algo}
    In Llama3 training~\cite{llama3}, the all-gather based {\tt pass-KV} algorithm is utilized, which initially performs an all-gather on the key and value tensors, followed by computing the attention output for the local query tensor chunk. The all-gather communication latency becomes a bottleneck in the critical path, complicating the overlap of operations during inference, especially with variant sequence lengths in a batch and partial prefill used in multi-turn chat. Conversely, the ring-based {\tt pass-KV} approach, while reducing the computation in smaller granularity, facilitates the overlapping of \textit{SendRecv} with attention computations within the ring loop.

    We further make a modification to the ring {\tt pass-KV} algorithm~\cite{liu2023ring} to better suit the partial prefill use case in multi-turn chats. Here an invariant we need to maintain for the ring algorithm is passing equal-sized messages between CP ranks to adhere to collective communication interfaces. CP ranks hold different numbers of KV embeddings as a result of multi-turn chat. Padding and decoding introduce slight variations in KV embedding length per rank even though our load-balanced sharding distributes KV embeddings evenly.

\begin{figure}[t]
    \centering
        \includegraphics[width=8cm]{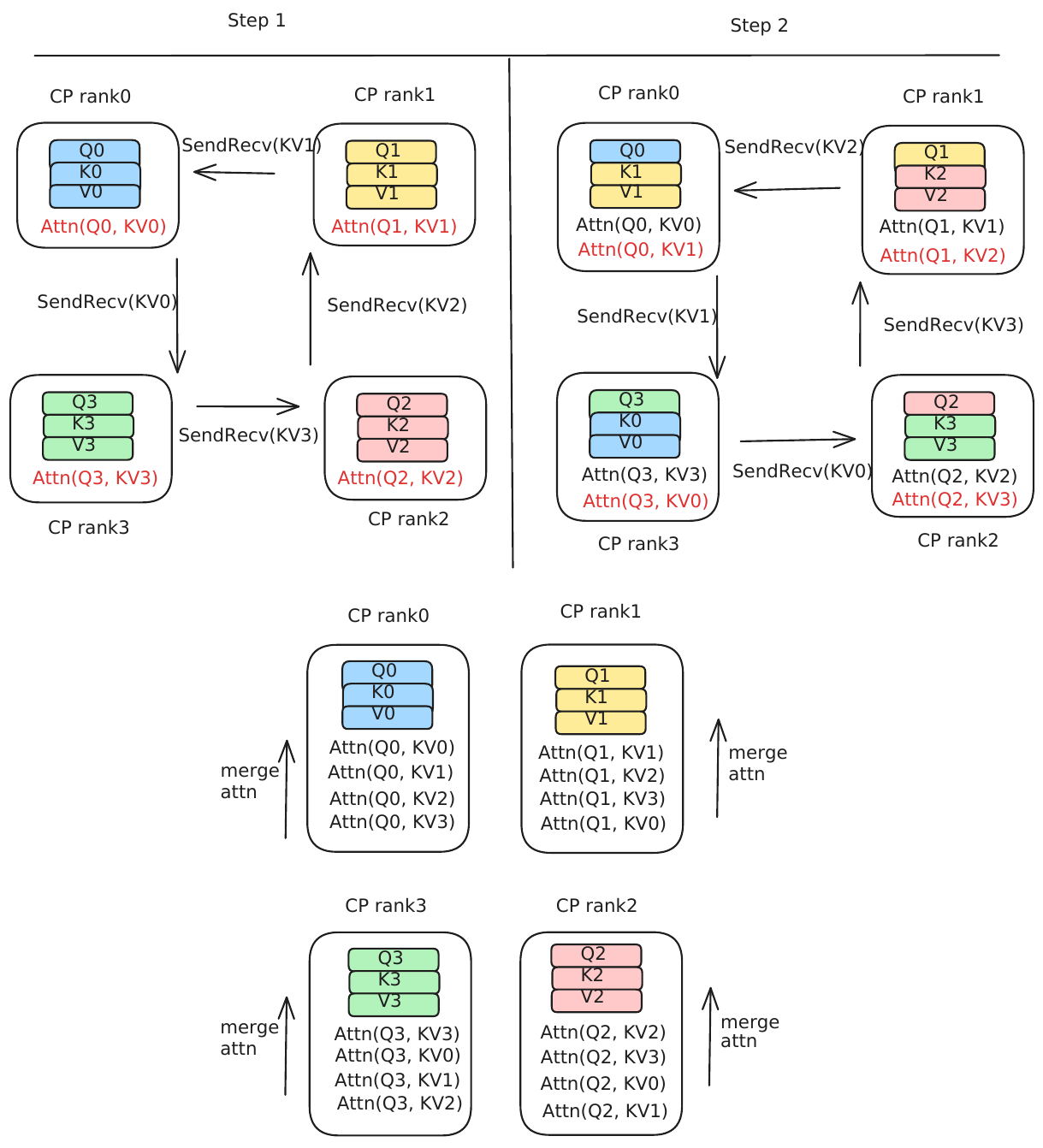}
        \caption{Ring Pass-KV Attention with 4 CP ranks (CP4).}
    \label{fig:pass-kv}
    \end{figure}

    Assume we have $ N $ CP ranks $ CP_0, CP_1, ..., CP_{N-1} $ with cached KV lengths of $ P_0, ..., P_{N-1} $, and partial prefill new tokens of length $T$. We pass KV embeddings of length $ \max_{0\le i< N}(P_i) + \lceil T / N \rceil $ around CP ranks in a ring (Figure~\ref{fig:pass-kv}), where $ \lceil T / N\rceil $ indicates the lengths of load-balanced sharding (Section \ref{section:load-balanced-sharding}) of $ T $ tokens over $ N $ ranks.

    \begin{algorithm}[tb]
       \caption{Fused Varseq Ring Pass-KV Partial Prefill}
       \label{alg:fused-varseq-ring-kv-prefill}
    \begin{algorithmic}
       \FOR{$i=0$ {\bfseries to} $ B-1 $}
       \STATE $L^i \leftarrow max_{0\le j < N}(P^{i}_{j} + T^{i}_{j})$
       \ENDFOR
       \STATE // On CP rank $ k $
        \STATE $KV^k_k \leftarrow concat_{i=0}^{B-1}(pad(P^{i}_{k} + T^{i}_{k}, L^i))$
        \STATE $Q_k \leftarrow concat_{i=0}^{B-1}(T^{i}_{k}) $
        \STATE $ p \leftarrow (k - 1) \mod N$
       \FOR{$ j=0 $ {\bfseries to} $ N-1 $}
       \STATE $ s   \leftarrow  (k -j) \mod N $
       \STATE Rank $k$ sends $KV^s_k$ to next rank
        \STATE Rank $k$ receives $KV^s_p$ from previous rank
       \STATE Compute $O_k^s \leftarrow GQA(Q_k, KV^s_k)$
        \STATE $KV^s_k \leftarrow KV^s_p$
       \ENDFOR
        \STATE Compute $O_k \leftarrow merge_{s=0}^{N-1}(O_k^s)$
    \end{algorithmic}
    \end{algorithm}

    For fused variable sequence lengths (Varseq) partial prefill of $ B $ sequences in one batch, assume we have sequences $S^0(P^{0}, T^{0}), ..., S^{B-1}(P^{B-1}, T^{B-1})$. The $ i $-th sequence $S^i$ has $P^{i}$ cached KV embeddings, $T^{i}$ new prefill tokens, with $P^{i}_{j}$ cached tokens and $T^i_j$ new tokens sharded to CP rank $j$. We have Algorithm~\ref{alg:fused-varseq-ring-kv-prefill} for a ring {\tt pass-KV} partial prefill with fused inputs for CP over $ N $ hosts. $KV^s_{k}$ indicates key and value embeddings received from rank $ k $ which is originally allocated to rank $ s $.

    In the ring algorithm, $ Q_k $, Q embeddings sharded to rank $ k $, need to attend to all key and value embeddings sharded to all ranks: $KV_0, KV_1, ..., KV_{N-1}$. The attention compute between $Q_k$ and $KV_j$ is overlapped with \textit{SendRecv} for $KV_{j-1}$ from a neighbor rank. We pass $KV_j$ in a ring $ N - 1 $ times and each rank executes $ N $ partial attention compute.

    At the end of the ring algorithm loop, each CP rank $ k $ will have the attention output of $ O_k^s $ with $ s = 0, 1, ..., N-1 $, where $ O_k^s $ denotes the attention output from $ Q_k $ and $ KV^s $ (key and value embeddings originally sharded to rank $ s $, see bottom of Figure \ref{fig:pass-kv}). We then apply a merge attention operator~\cite{juravsky2024hydragen} to get the result of $ Q_k $ interacted with all $ KV $ embeddings across CP ranks (See Appendix \ref{sec:merge-attn}, Equation \eqref{eqn:merge_attn}).

    \subsubsection{Ring Pass-Q Algorithm}
    \label{sec:ring-pass-q-algo}
\begin{figure}[t]
    \centering
    \includegraphics[width=8cm]{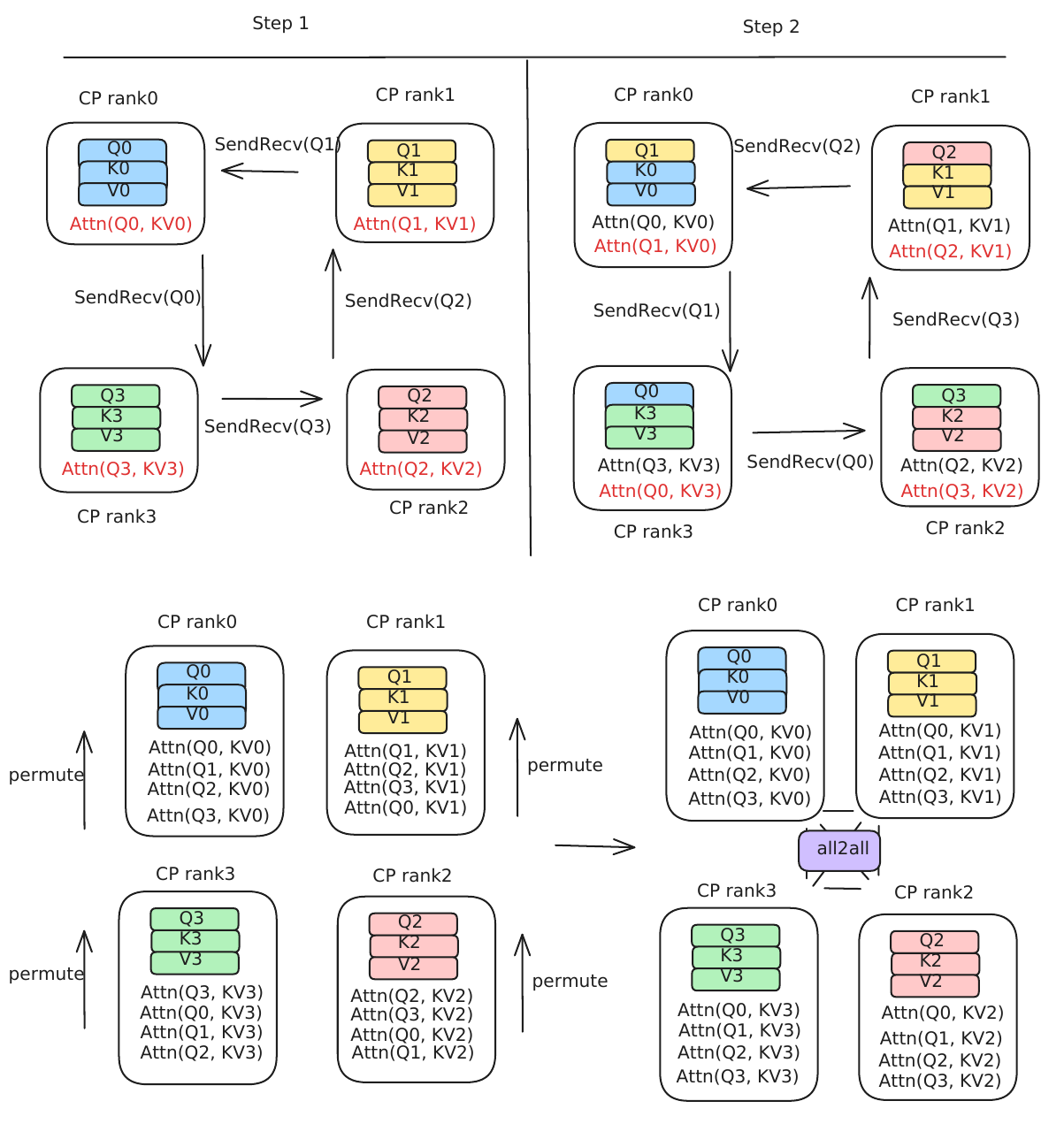}
    \caption{Ring Pass-Q Attention with 4 CP ranks (CP4).}
    \label{fig:pass-q}
    \end{figure}

    Passing Q embeddings around while keeping K and V embeddings stationary will have partial attention results scattered across CP ranks. We need to have another round of collective communication over CP process group to restore the partial outputs to the original source rank. Following the notations of ring {\tt pass-KV} algorithm in Section \ref{sec:ring-pass-kv-algo}, we have Algorithm \ref{alg:fused-varseq-ring-q-prefill} for ring {\tt pass-Q} attention (Figure~\ref{fig:pass-q}). Similarly, $Q^s_{k}$ indicates a Q embedding received from rank $k$ which was initially allocated to rank $s$. Note that with {\tt pass-Q} we have the guarantee that all CP ranks have the same embedding lengths for query as a result of load-balanced sharding (Section \ref{section:load-balanced-sharding}).

    \begin{algorithm}[t]
       \caption{Fused Varseq Ring Pass-Q Partial Prefill}
       \label{alg:fused-varseq-ring-q-prefill}
    \begin{algorithmic}
       \STATE // On CP rank $k$ with $KV_k$
        \STATE $Q_k \leftarrow concat_{i=0}^{B-1}(T^{i}_{k}) $
        \STATE $ p \leftarrow (k - 1) \mod N$
       \FOR{$ j=0 $ {\bfseries to} $ N-1 $}
       \STATE $ s   \leftarrow  (k -j) \mod N $
       \STATE Rank $k$ sends $Q^s_k$ to next rank
        \STATE Rank $k$ receives $Q^s_p$ from previous rank
       \STATE Compute $O_s^k \leftarrow GQA(Q^s_k, KV_k)$
        \STATE $Q^s_k \leftarrow Q^s_p$
       \ENDFOR
        \STATE Permute $\{O_s^k\}_{s=0}^{N-1}$ and \textit{All2All} to recover $\{O_k^s\}_{s=0}^{N-1}$.
        \STATE Compute $O_k \leftarrow merge_{s=0}^{N-1}(O_k^s)$
    \end{algorithmic}
    \end{algorithm}

      \begin{algorithm}[tb]
       \caption{Batched Ring Pass-Q Decode}
       \label{alg:ring-q-decode}
    \begin{algorithmic}
       \STATE // On CP rank $k$ with $KV_k$, query $Q_k$, batch ids $bid_k$
        \STATE $ p \leftarrow (k - 1) \mod N$
       \FOR{$j=0$ {\bfseries to} $ N-1 $}
       \STATE $ s   \leftarrow  (k -j) \mod N $
       \STATE Rank $k$ sends $Q^s_k$, $bid^s_k$ to next rank
        \STATE Rank $k$ receives $Q^s_p$, $bid^s_p$ from previous rank
       \STATE Compute $O_s^k \leftarrow GQA(Q^s_k, KV_k[bid^s_k])$
        \STATE $Q^s_k \leftarrow Q^s_p$
        \STATE $bid^s_k \leftarrow bid^s_p$
       \ENDFOR
        \STATE Permute $\{O_s^k\}_{s=0}^{N-1}$ and \textit{All2All} to recover $\{O_k^s\}_{s=0}^{N-1}$.
        \STATE Compute $O_k \leftarrow merge_{s=0}^{N-1}(O_k^s)$
    \end{algorithmic}
    \end{algorithm}

    \textit{All2All} for partial attention outputs is on the critical path and therefore introduces an additional communication overhead apart from the communication for passing query embedding. The analysis for overlapping query embedding and attention in Equation \eqref{eqn:pass-kv-overlap} and \eqref{eqn:pass-q-overlap} only applies to the ring communication. The heuristics in Algorithm \ref{alg:pass-kv-vs-pass-q-partial} for switching between \passkv{} and \passq{} doesn't take \textit{All2All} latency into account\footnote
{We present a refined algorithm in Appendix \ref{sec:all2all} and provide a detailed time breakdown for validations in Table \ref{tab:passKV_passQ_breakdown_all}.}.

\subsection{Ring Pass-Q Decode}
\label{sec:ring-pass-q-decode}
    With multi-turn prefill and decode, key and value embeddings of the decode tokens are also stored in the KV cache. As decoding generates one response token at a time for each sequence, each decode batch contains exactly one token for each sequence in the batch. If context-parallel decode consistently shards the decoding tokens of a sequence to one specific rank, the rank that handles both decode and prefill will encounter load imbalance issues: it will have longest KV cache and out-of-memory (OOM) before other ranks reach their KV cache capacity.

    To ensure we utilize full KV cache capacity from all CP ranks, we implemented \textit{batched ring pass-Q decode} where we offset by 1 index for each decode iterations and shard batched decode evenly with round-robin. With exactly 1 token per sequence for decode, we pass Q rather than K and V embeddings to minimize communication size (Equation \ref{eqn:pass-q-vs-kv}). Algorithm \ref{alg:ring-q-decode} summarizes our CP decode algorithm with the same notations used for prefill algorithms.

    Similar to ring {\tt pass-Q} prefill, we need to permute the partial attention output order and communicate scattered partial attention outputs back to the original source ranks.

\section{Experiments}
\subsection{Experiment Setup}
\label{sec:exp_setup}
    We used Llama3 405B model with row-wise quantized FP8 weights~\cite{llama3} for feed forward layers after GQA. Llama3 405B is a dense transformer model with 126 transformer layers, 16384 model dimension, 128 query heads, and 8 key and value heads (Table \ref{tab:llama3_model_config}).

    We ran our performance benchmarks on the Grand Teton platform \cite{gt2022}, where each host has 8 Nvidia H100 GPUs fully connected with NVLink (``host'' and ``node'' are interchangeable in the subsequent text).
    Each H100 GPU is equipped with 96GB HBM2e with 2.4 TB/sec peak memory bandwidth.
    We tested on two subtypes of Grand Teton platforms: Grand Teton Training (GTT) and Grand Teton Inference (GTI). GTT hosts are inter-connected with backend RDMA network with 400 Gb/s per GPU, and GTI hosts are inter-connected with frontend network over TCP/IP with 100 Gb/s per GPU.

    With row-wise FP8 quantization\footnote{\url{https://github.com/pytorch/FBGEMM/tree/main/fbgemm_gpu/experimental/gen_ai}}, the entire 405B model fits into one node with TP8 (tensor parallelism across 8 partitions) partitioning. Each GPU holds 1 KV head and 16 Q heads, and feed forward layers are partitioned with alternating column and row parallelism~\cite{shoeybi2019megatron}. Flash Attention 3~\cite{shah2024flashattention} is adopted for attention kernels in prefill, while Flash Decoding \cite{flash_decoding} with number of K/V splits 256 is used during decoding.

\begin{figure}[t]
    \centering
    \includegraphics[width=8cm]{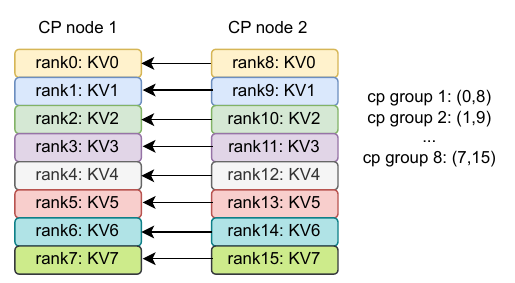}
    \caption{Context parallel across nodes and tensor parallel within nodes, with 2 CP ranks (CP2).}
    \label{fig:cp-tp-8-way-p2p}
    \vspace{-0.2in}
    \end{figure}

    We tested full prefill, partial prefill, and decode performance with context parallelism over 1-16 nodes. Within each CP node the model is partitioned with TP8 over 8 GPUs. We form one CP communication group per KV head, with each CP group consisting of $ N $ GPUs (one GPU in each node) holding the same KV head in their respective tensor parallel groups. Ring communication around CP ranks is implemented an 8-way \textit{SendRecv} (Figure \ref{fig:cp-tp-8-way-p2p}).

 \subsection{Context Parallel Prefill Scaling}
     \subsubsection{Latency Reduction with Fixed Context Length}
     Llama3 405B model supports a maximum of 128K context window, which is equivalent to 300-400 pages of books. We used max batch size 1 and tested how the full prefill latency for context lengths 2K to 128K vary with respect to the addition of more CP nodes.
 
\begin{figure}[t]
        \subfigure[GTT Latency for CP with 1, 2, 4, 8 nodes.]{
            \includegraphics[width=8cm]{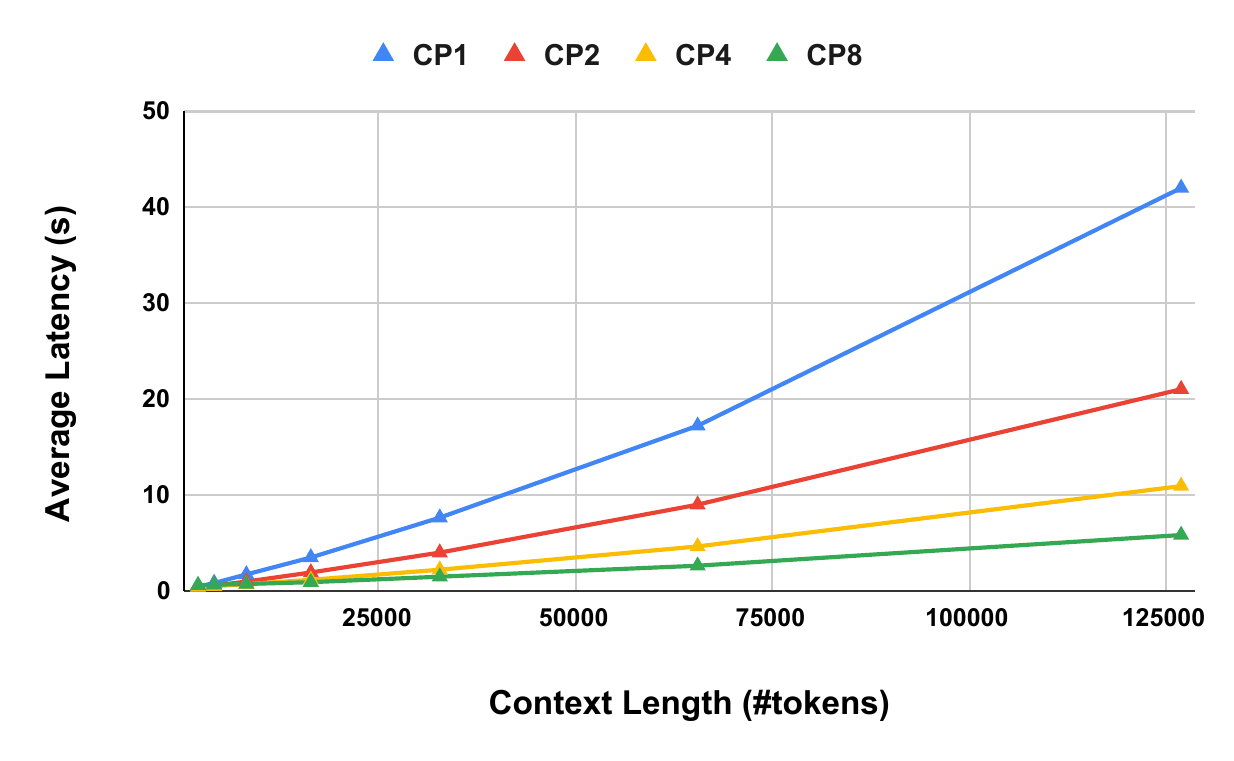}
            \label{fig:405b-full-prefill-pass-kv}
        }

        \subfigure[GTI Latency for CP with 1, 2, 4 nodes]{
            \includegraphics[width=8cm]{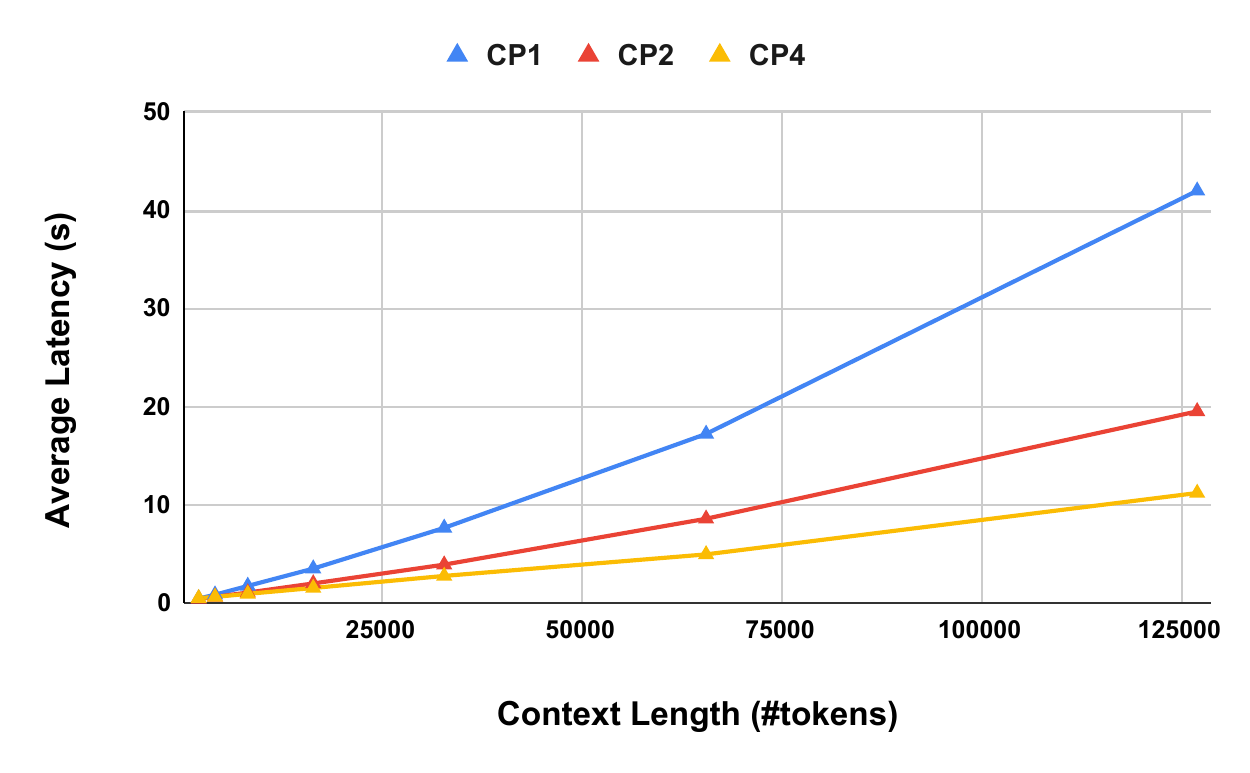}
            \label{fig:405b-full-prefill-pass-kv-gti}
        }
        \caption{Llama3 405B {\tt pass-KV} full prefill latency.}
    \end{figure}

     Figure \ref{fig:405b-full-prefill-pass-kv} shows the full prefill latency of {\tt pass-KV} full prefill on GTI and GTT for 1-8 CP nodes. With sufficiently large context lengths, the latency for passing key and value embeddings are overlapped with attention compute, and we get proportional latency reduction with more CP nodes: latency for the same input length is halved as we double the number of CP nodes. Specifically, with CP8 on GTT, an FP8 Llama3 405B model can process a 128K token prefill in 5.85 seconds.
 
     For GTI systems with much lower inter-host bandwidth over frontend TCP/IP network, we observe the same scalability with up to 4 nodes. Inspecting the GPU trace from GTI, we found the achieved bandwidth for inter-host communication is roughly 3GB/s per rank, which is still enough to overlap the {\tt pass-KV} communication with attention compute, demonstrating the robustness of {\tt pass-KV} algorithm even with low inter-connect bandwidth.

\subsubsection{Comparing with Multi-Node Tensor-Parallel}
    \label{sec:cp-vs-tp}
    To compare with context-parallel performance, we benchmarked tensor-parallel over multiple nodes on GTT with up to 8 nodes. Llama3 405B model has 8 KV heads. To effectively parallelize 8 KV heads across more than 8 GPUs, we replicate each KV head over $ N_{TP} / N_{KV} $ GPUs where $ N_{TP} $ is the total number of GPUs in the tensor parallel group and $N_{KV}$ is the number of KV heads. Query heads are distributed evenly to all GPUs with $ N_{H} / N_{TP} $ query heads per GPU. Computation is still fully parallelized over $ N_{TP} $ GPUs.

    We calculate scaling ratio for a paralellization across $ N $ nodes as as $ \tau_1 / \tau_N $, where $ \tau_N $ is the latency for $ N $ nodes to process a 128K context prefill. Better parallelization algorithms would have scaling ratios closer to $ N $.

\begin{figure}[h]
    \centering
    \includegraphics[width=8cm]{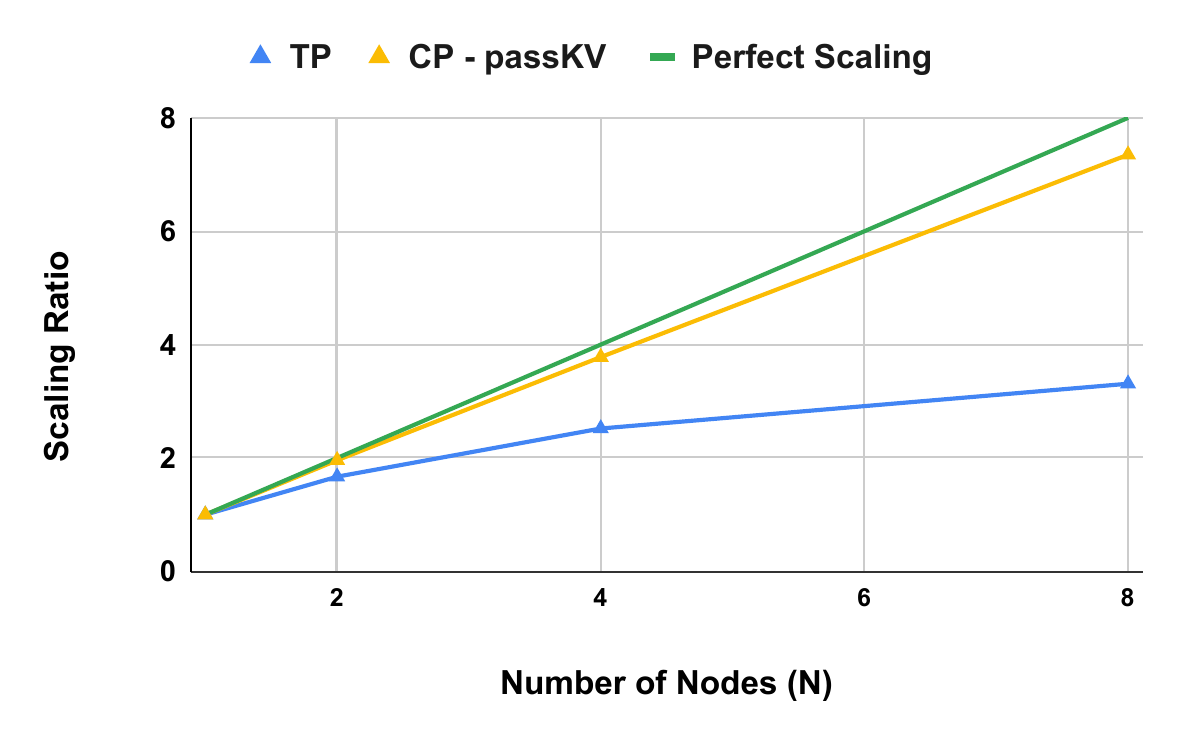}
    \caption{Scaling ratio (latency with one node over latency with N nodes) of context parallel vs. multi-node tensor parallel.}
    \label{fig:scaling-ratio-cp-tp}
    \end{figure}

Figure \ref{fig:scaling-ratio-cp-tp} illustrates the scaling ratios for multi-node tensor parallelism compared to context parallelism across 1 to 8 GTT nodes.
    Tensor-parallel becomes more bottlenecked by inter-host communication with the growth of capacity,
as \textit{AllReduce} latency increased significantly with the addition of more nodes.
While the latency is different by roughly 15\% between CP2 and TP16 on 2 nodes, the difference drastically increases to 100\% when scaled to 8 nodes.

    This evaluation is performed on H100 hosts which exhibit significantly lower inter-host bandwidth compared to intra-host badwidth. For future GB200~\cite{nvgb200} with NVLink connecting multiple hosts, tensor parallelism can still benefits with reasonable scalability.

    \subsubsection{Scaling Context Length with Fixed Capacity}

    \begin{figure}[h]
    \centering
    \includegraphics[width=8cm]{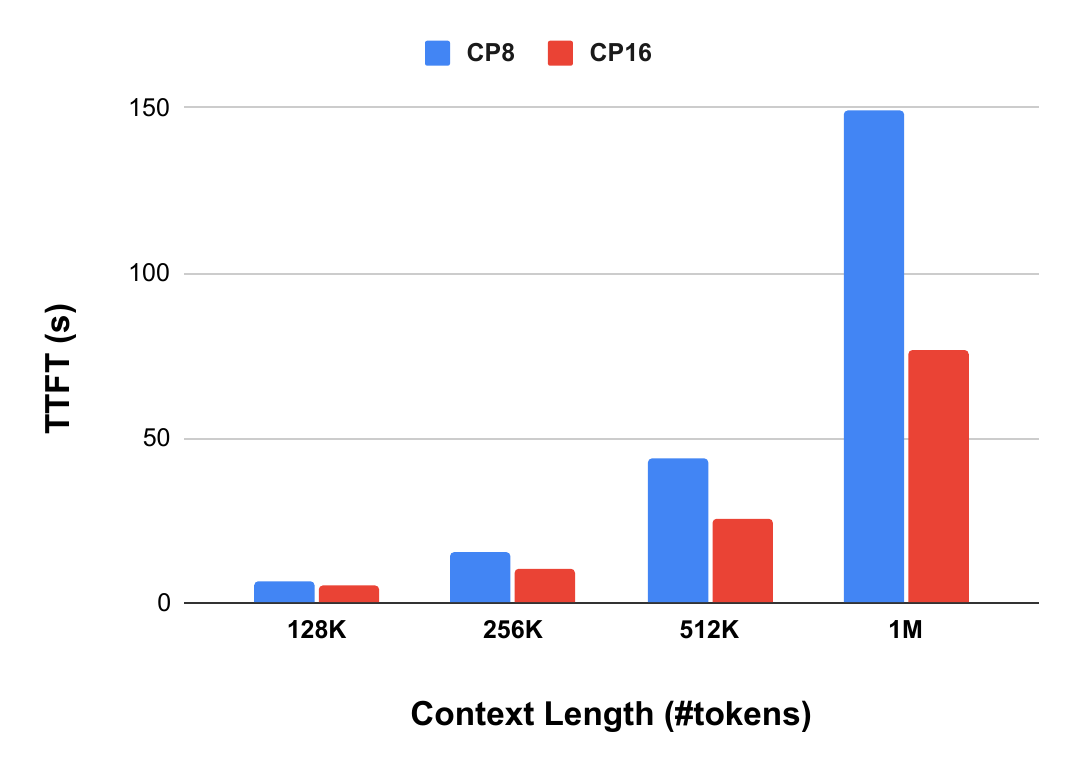}
        \caption{TTFT of 128K-1M context with 8 and 16 CP ranks (CP8 and CP16).}
    \label{fig:ttft-1m}
    \end{figure}

    By partitioning the KV cache across CP ranks, we also enhance the KV cache capacity as more CP nodes are added. To demonstrate scalability in terms of both capacity and latency, we run up to 1M context prefill over 8 and 16 GTT nodes. This corresponds to approximately 1 hour of video content.
    With a 16-node setup, we achieve an exact prefill in 77 seconds for a 1M context length and 3.8 seconds for a 128K context length (Figure \ref{fig:ttft-1m}).
    The quadratic increase in attention latency with context length begins to dominate the overall time to first token (TTFT) latency, resulting in more than $ 2 \times $ increase in TTFT with a $ 2 \times $ increase in context length for $ \geq $ 512K token prefill.

    We calculate the FLOPS utilization of a 1M context length on 16 nodes in Appendix \ref{sec:mfu}. The achieved FLOPS is 502 TF/sec per H100, compared to a standalone Flash Attention v3 benchmark performance of 540 TF/sec for 8K context length (1M over 128 H100 GPUs) on a single GPU, resulting in a 93\% parallelization efficiency. Considering the peak FLOPS on the specialized H100 configurations, we achieve approximately 63\% FLOPS utilization.

    \subsubsection{Pass-KV vs. Pass-Q Partial (Persistent KV) Prefill}

    \begin{table}[t]
        \caption{TTFT (in $ ms $) for {\tt pass-KV} vs. {\tt pass-Q} varying $ P $ and $ T $ with $ P+T=128000 $, on 4 CP ranks (CP4). $ P $: length of existing tokens in the KV cache, $ T $: length of new tokens.
}
    \vskip 0.05in
    \begin{center}
    \begin{small}
    \begin{sc}
    \begin{tabular}{c | c | c | c | c}
    \whline
        $ P $ & $ T $ & Miss Rate & {\tt pass-KV} & {\tt pass-Q} \\
    \hline
126720 & 1280 & 1.00\% & 1023.39 & 898.71 \\ \hline
124800 & 3200 & 2.50\% & 1110.18 & 1046.43 \\ \hline
\textbf{123840} & \textbf{4160} & \textbf{3.25\%} & \textbf{1298.92} & \textbf{1280.1} \\ \hline
\textbf{121600} & \textbf{6400} & \textbf{5.00\%} & \textbf{1305.56} & \textbf{1302.01} \\ \hline
115200 & 12800 & 10.00\% & 2080.67 & 2205.27 \\ \hline
102400 & 25600 & 20.00\% & 3353.02 & 3617.02 \\ \hline
89600 & 38400 & 30.00\% & 4629.23 & 4922.52 \\ \hline
76800 & 51200 & 40.00\% & 5745.08 & 6217.83 \\ \hline
64000 & 64000 & 50.00\% & 6845.21 & 7367.99 \\ \hline
51200 & 76800 & 60.00\% & 7890.35 & 8468.66 \\ \hline
38400 & 89600 & 70.00\% & 8697.27 & 9666.62 \\ \hline
25600 & 102400 & 80.00\% & 10105.78 & 10652.39 \\ \hline
12800 & 115200 & 90.00\% & 11136.4 & 11571.62 \\ \hline
0 & 128000 & 100.00\% & 11462.15 & 12360.57 \\
    \whline
    \end{tabular}
    \end{sc}
    \end{small}
    \end{center}
    \label{tab:passKV_passQ_comp}
    \vskip -0.1in
    \end{table}

   The persistent KV cache provides substantial advantages in long-context LLM inference by minimizing repeated computational overhead in multi-turn conversations.
   In Table \ref{tab:passKV_passQ_comp}, experiments with a 128K context length on 4 GTT nodes demonstrated that, in both {\tt pass-KV} and {\tt pass-Q} implementations, TTFT latency is linearly proportional to the persistent \emph{KV cache miss rate} (${T\over T+P}$).


    \begin{figure}[t]
    \centering
    \includegraphics[width=8cm]{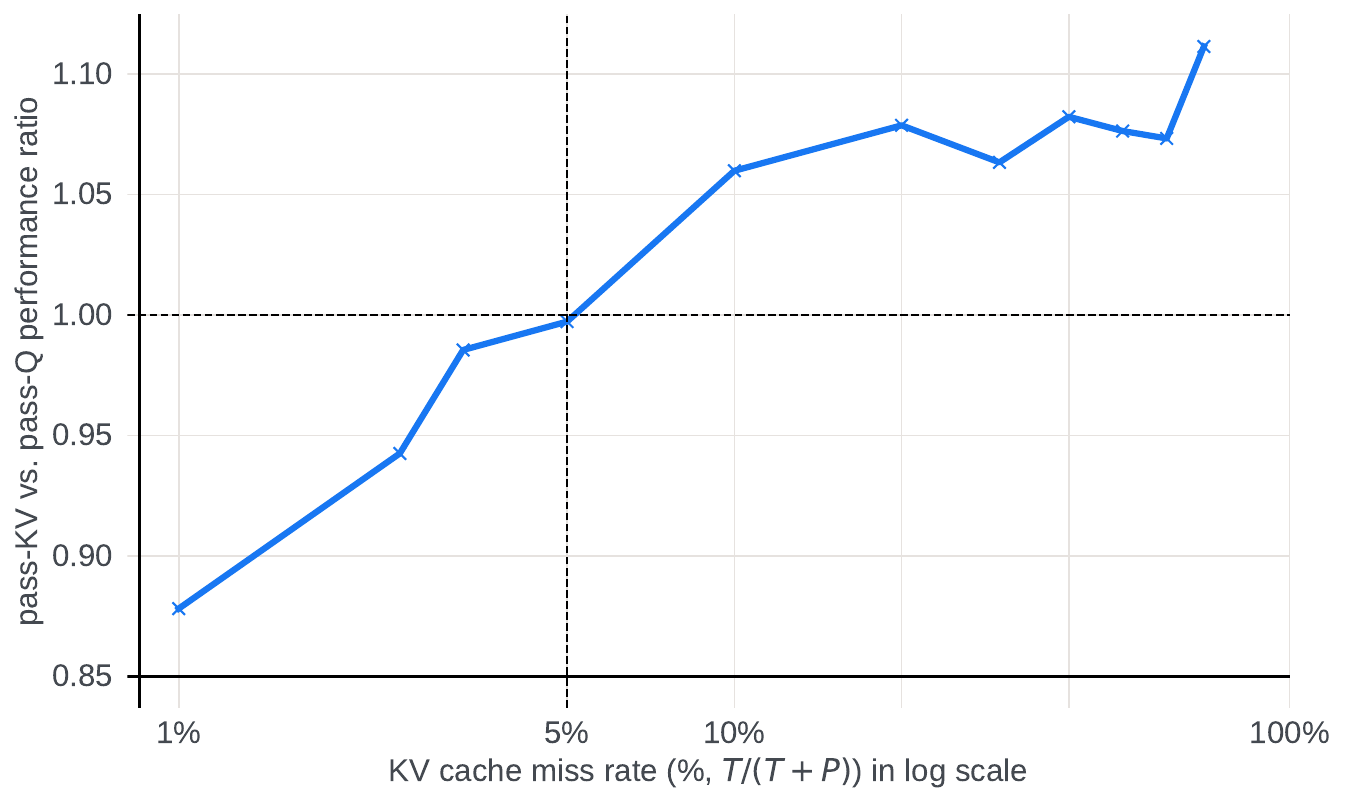}
        \caption{{\tt pass-KV} / {\tt pass-Q} speed ratio of 128K context with persistent KV cache miss rate, varying $ P $ and $ T $ with $ P+T=128000 $, on 4 CP ranks (CP4).}
    \label{fig:pass-kv-q-ratio-pkv}
    \end{figure}

Figure \ref{fig:pass-kv-q-ratio-pkv} compares {\tt pass-KV} and {\tt pass-Q} in terms of the KV cache miss rate. When the KV cache miss rate is less than 5\%, {\tt pass-Q} exhibits better latency; however, when the miss rate exceeds 5\%, {\tt pass-KV} achieves lower latency.

The tipping point between {\tt pass-Q} and {\tt pass-KV} occurs at $ T=6400 $ (5\% KV cache miss rate).
Table \ref{tab:passKV_passQ_breakdown_all} details the time breakdown for cache miss rates slightly below and above this configuration (2.5\% and 10\% miss rate).
\textit{SendRecv} and {\sc{Attn}} represent the \textit{SendRecv} time and the partial attention compute time (in $\mu s$) for each iteration of the ring algorithm loop, which is repeated $ N -1 $ times. The \textit{All2All} time refers to the communication required in the merge attention step at the end of {\tt pass-Q} algorithm. Note that for $ T = 3200 $, the sum of exposed {\tt pass-KV} communication ($ (N-1) \cdot $ (\textit{SendRecv} - {{\sc{Attn}}})) is longer than {\tt pass-Q} \textit{All2All}, resulting in better performance for {\tt pass-Q} compared to {\tt pass-KV}.

We further validate the analytical model in Algorithm \ref{alg:pass-kv-vs-pass-q-partial} for predicting the selection of {\tt pass-KV} vs. {\tt pass-Q} from Table \ref{tab:passKV_passQ_comp}.
\begin{itemize}
    \item When the KV cache miss rate exceeds 12.5\% ($ = 2 \cdot \frac{N_{KV}}{N_H} $ in Equation \ref{eqn:pass-q-vs-kv}), {\tt pass-KV} is always selected, meeting the 2nd condition in Algorithm \ref{alg:pass-kv-vs-pass-q-partial}.
    \item At 10\% KV cache miss rate, {\tt pass-KV} remains the choice since the number of new tokens $T$ is sufficiently large, satisfying Equation \ref{eqn:pass-kv-overlap} (with \textit{SendRecv} hidden under {\sc{Attn}} in Table \ref{tab:passKV_passQ_breakdown_all}).
    \item Around 5\% cache miss rate (e.g., $ T=6400 $), the differences between \passkv{} and \passq{} is less than 1\%, allowing for either option to be selected.
    \item When cache miss rate falls below 3.25\%, {\tt pass-KV} communication becomes exposed, leading to the selection of {\tt pass-Q}. Specifically, at a 2.5\% cache miss rate, the sum of the exposed communication in {\tt pass-KV} ring loop is larger than \textit{All2All} exposed in {\tt pass-Q} (Equation \ref{eqn:pass-kv-le-pass-q-a2a-1}, Appendix \ref{sec:all2all}), resulting in the selection of \passq{}.
\end{itemize}

    \begin{table}[t]
        \caption{Time breakdown (in $\mu s$) on {\tt pass-KV} vs. {\tt pass-Q} ring attention at cache miss rate of $ 2.5\% $ and $ 10\% $ with $ P+T=128000 $, on 4 CP ranks (CP4).}
    \vskip 0.15in
    \begin{center}
    \begin{footnotesize}
    \setlength{\tabcolsep}{4pt}
    \begin{tabular}{ c | r | ccc}
    \whline
        Miss Rate & {\tt pass-KV/Q} & \textit{SendRecv} & {\sc{Attn}} & \textit{All2All} \\
        \hline
        \multirow{2}{*}{ $ 2.5\% $} & \passkv{} & 627 & 414 & N/A \\
        & \passq{} & 166 & 414 & 424 \\
        \hline
        \multirow{2}{*}{ $ 10\% $} & \passkv{} & 631 & 1608 & N/A \\
        & \passq{} & 544 & 1608 & 1023 \\
    \whline
    \end{tabular}
    \end{footnotesize}
    \end{center}
    \label{tab:passKV_passQ_breakdown_all}
    \vskip -0.1in
    \end{table}

%
%

\subsection{Decode Performance}
    Inference decode generates one output token at a time, resulting in a small amount of computation workloads and communication traffic. To avoid host kernel launch bottlenecks for these small kernels, we run both CP and TP decode with CUDA Graphs \cite{cudagraph}.

    \textbf{Context Length Scalability:} We benchmarked CP decoding performance with 2 nodes on GTT (using ring {\tt pass-Q} decode algorithm in Section \ref{sec:ring-pass-q-decode}), and compare with TP8 decoding performance on 1 node using a single batch decode with various context lengths.
    As shown in Table~\ref{tab:ttit_tp8_cp2}, the TTIT of both TP8 and CP2 does not increase too much: For both TP8 and CP2, the computation and communication for linear layers stay the same while the latency of attention kernels increases with a longer context length.

\begin{table}[t]
\centering
    \caption{TTFT / TTIT (in $ ms $) comparisons between TP8 and CP2 with different context lengths at batch size 1.}
    \vskip 0.15in
{\footnotesize
    \begin{tabular}{c | cc | cc}
        \whline
        \multirow{2}{*}{Context length} & \multicolumn{2}{c | }{ TP8 } & \multicolumn{2}{c}{ CP2+TP8 } \\
        & TTFT & TTIT & TTFT & TTIT \\
        \hline
        8K  & 1740  & 44.51 & 999  & 65.61 \\
        32K & 7658  & 44.64 & 4015  & 65.66 \\
        128K & 42010 & 46.26 & 21042 & 66.63 \\
        \whline
    \end{tabular}
}
\label{tab:ttit_tp8_cp2}
\end{table}

\begin{table}[t]
\centering
    \caption{TTFT / TTIT (in $ ms $) comparisons between TP8, CP2, TP16, CP4, TP32 with 128K context length at batch size 1.}
    \vskip 0.15in
{\footnotesize
\setlength{\tabcolsep}{20pt}
\begin{tabular}{c|c|c}
\whline
 & TTFT & TTIT \\ \hline
CP1+TP8             & 42010            & 46.26            \\ \hline
CP2+TP8    & 21042            & 60.23            \\ 
TP16              & 29917            & 39.52            \\ \hline
CP4+TP8   & 10950            & 71.31            \\ 
TP32             & 19841            & 47.3             \\ \hline
\whline
\end{tabular}
}
\label{tab:ttit}
\vskip -0.15in
\end{table}


    \textbf{Parallelism Scalability:} We benchmarked different parallelization configurations up to four CP nodes to observe the scalability of both TP and CP.
    Table~\ref{tab:ttit} shows that TTIT tends to be longer for both scaling TP and scaling CP.
    TTIT for scaling TP increases to 47 $ ms $ while TTIT for scaling CP increases to 71 $ ms $.
    Both TP and CP have poor scalability for decoding when adding more hosts (e.g., using 4 nodes can result in worse TTIT than using a single node). For TP, lower computation latency on linear layers is offset by increased communication latency increased.
    
    For CP, as we increase the number of hosts, the effective length seen by each attention kernel decreases, so each individual attention op becomes faster (Table ~\ref{tab:attn_scaling}). However TTIT still degrates compared to CP=1, and the reason for that is two-fold: (1) Current implementation pads the number of queries to make it divisible by the number of ranks, which for B=1 means the total number of processed queries increases with CP. (2) The communication latency - sending Q chunks to the next rank at each iteration of the loop and \textit{All2All}-exchanging partial attention outputs after the loop - also grows with the number of hosts. As a result, the total {\tt pass-Q} attention latency and TTIT increase with CP.

\begin{table}[t]
\centering
    \caption{Attention scaling with the number of CP hosts (Time in $ \mu s$).}
    \vskip 0.15in
{\footnotesize
   \begin{tabular}{c|c|c|c}
        \whline \\
        & \multicolumn{3}{c}{Context length 128K, batch size 1} \\
          \\
         & TP8 & CP2+TP8 & CP4+TP8 \\
        \hline
        Effective context length  &  128K & 64K & 32K  \\
        Individual attention op &  38.9 & 22.0 & 14.7  \\
        Attn (whole ring loop) & 38.9 & 43.2 & 60.8 \\
        \textit{SendRecv} & 0 & 32.3 & 105.7 \\
        \textit{All2All} & 0 & 81.1 & 79.9 \\ 
        Whole {\tt pass-Q} & 38.9 & 157.7 & 238.6 \\
        \hline \\
        & \multicolumn{3}{c}{Context length 32K, batch size 4} \\
        \\
        \hline
        Effective context length  &  32K & 16K & 8K  \\
        Individual attention op &  60.1 & 13.9 & 9.6  \\
        Attn (whole ring loop) & 60.1 & 24.5 & 41.3 \\
        \textit{SendRecv} & 0 & 33.3 & 104.9 \\
        \textit{All2All} & 0 & 66.8 & 72.2 \\ 
        Whole {\tt pass-Q} & 60.1 & 136.6 & 180.6 \\
        \whline
    \end{tabular}
}
\label{tab:attn_scaling}
\end{table}

    In summary, context parallel is best suited for improving prefill performance and can be best leveraged with a serving system that decouples the parallelization scheme for prefill and decode~\cite{qin2407mooncake,zhong2024distserve}.
    For standalone deployment where prefill and decode are both on the same set of hosts, CP drastically improves the prefill latency, at the expense of decode latency regression (Removing batch padding and better overlap of computation and communication can help to minimize this regression). 

\section{Conclusion}
In conclusion, our work highlights the effectiveness of context parallelism and ring attention variants in improving the efficiency of LLM inference for long-context scenarios. By leveraging up to 128 GPUs, we achieved near-linear scaling and significantly reduced latency, completing tasks with impressive speed and efficiency.
Our implementation of the lossless exact {\tt pass-KV} and {\tt pass-Q} ring attention variants has been critical in supporting various full prefill, partial prefill, and decoding scenarios. The runtime heuristic adaptively selects {\tt pass-KV} or {\tt pass-Q} based on KV cache hit rate,  optimizing their application for the most suitable scenarios.

As we keep improving LLM's capacity to understand increasingly longer and more complex  context, one can expect diminishing utility with exact attention over all historical tokens. More efficient algorithms for retrieving a small subset of information from a much larger context to answer simple probe questions will be increasingly important. While context parallel is an efficient exact algorithm for scaling exact attention with more capacity, combining its processing power with an approximate retrieval algorithm for ultra-long context may be the best way to bound the processing latency for context window growth at and beyond 1M.

\section{Acknowledgments}

We express our gratitude to our outstanding colleagues for their significant contributions to various aspects of LLM inference. Special thanks go to Geonhwa Jeong, Jaewon Lee, Jason Park, Vlad Mihailescu, Zheng Yan, and Daniel Haziza for their invaluable efforts related to this paper.
Our thanks also go to Chunqiang Tang for his early feedback and proofreading of the draft.
Furthermore, we appreciate the leadership and support provided by Chunqiang Tang, Maxim Naumov, Amit Nagpal, Tony Liu, Changkyu Kim, Anca Agape, and Stephen Chen.

\clearpage
\newpage

\balance
\bibliographystyle{mlsys2025}
\bibliography{cp_cite}


\clearpage
\newpage
\appendix

\balance


\section{MFU Calculation for 1M context length}
\label{sec:mfu}
\begin{table}
\caption{Llama3 405B model configurations.}
\centering
{\footnotesize
\setlength{\tabcolsep}{20pt}
  \begin{tabular}{l | c }
  \whline
  Parameter & Value  \\
  \whline
Layers ($ \#layers $) &  126 \\
\hline
Model Dimension ($ D $) &  16,384 \\
\hline
FFN Dimension & 53,248 \\
\hline
Attention Heads ($ N_H $) & 128 \\
\hline
Key/Value Heads ($ N_{KV} $) & 8 \\
\hline
      Parameter Size ($ W $) & 405 B \\
  \whline
  \end{tabular}
}
\label{tab:llama3_model_config}
\end{table}
    We calculate the effective Model FLOPS utilization (MFU)~\cite{chowdhery2023palm} in this section. The Llama3 405B model configurations are listed in Table \ref{tab:llama3_model_config}. The total FLOPS are dominant by GEMM and Attention parts:
\[
    \textbf{\tt Total FLOPS} = \textbf{\tt GEMM FLOPS} + \textbf{\tt ATTN FLOPS}.
\]
    \begin{itemize}
        \item For GEMM, an $ W $-parameter Transformer model requires $ 2 \cdot W $ matrix multiplication FLOPs for each token during inference:
\[
    \textbf{\tt GEMM FLOPS} = 2 \times W \times T \times B.
\]
\item For Attention, the FLOPS is quadratic with respect to the context length $ T $:
\[
\textbf{\tt ATTN FLOPS} = 1/2 \times 4 \times B \times T^2 \times D \times \#layers,
\]
where 1/2 is from the causal mask, 4 is from 2 batch matmul and 2 FLOPS for multiplication and addition.
    \end{itemize}

With input sequence length $ T = 1M $, batch size $ B = 1 $, the parameter size
    $ W = 405B $, we can get \textbf{\tt GEMM FLOPS} = $ 2 \times 405B \times 1M $ = $ 8.1 \times 10^{17} $.
With the model dimension $ D = 16384 $, and number of layers $ \#layers = 126 $,
we can derive \textbf{\tt ATTN FLOPS} = $ 1/2 \times 1M ^ 2 \times 16384 \times 126 $ = $ 4.1 \times 10^{18} $.
Attention FLOPS is more dominant compared to GEMM FLOPS.
The total FLOPS is $ 4.9 \times 10^{18} $. With 77 seconds for 1M context length using 128 H100 GPUs, each H100 achieves
$ 4.9 \times 10^{18} / 77 / 128 = 502 $ TF/sec.
    Note that with the standalone Flash Attention v3 causal attention benchmark using 8K context length on a single H100 (1M context length sharded across 128 H100 GPUs), we achieve 540 TF/sec.
    One caveat for the evaluation is that GTT/GTI (Section \ref{sec:exp_setup}) are configured with power limited H100 GPUs (500 Watt) with lower memory bandwidth (96 GB HBM2e with 2.4 TB/sec instead of 80 GB HBM3 with 3.35 TB/sec), where the BF16 peak for each H100 is 800 TF/sec, instead of 989 TF/sec for H100 HBM3 with 700 Watt.





\section{Merge Attention}
\label{sec:merge-attn}
The idea of merging attention outputs from different keys/values originates from Online Softmax~\cite{milakov2018online}. Later this idea was reused in Flash Attention~\cite{fa_v1,fa_v2}. Here we derive the equation to merge the partial attention outputs from different CP ranks.

The scaled dot production attention operates on query/key/value tensors $ Q/K/V$. For simplicity, we don't consider various mask like causal masks (no batching or multiple attention heads either). There is one $Q/K/V$ corresponding to each sequence position. $Q/K/V$ at a given sequence position is a vector in the embedding space. The attention output is defined as
\[
O = {\tt Attn}(Q, K, V) = {\tt softmax}\left(\frac{Q K^T}{\sqrt{d}}\right) V,
\]
where {\tt softmax} is applied row-wise.

Assuming the size of row is $ R $,
\[
    O = \frac{\sum_{i=0}^{R-1} \text{\tt exp}^{Q \cdot K_i^T / \sqrt{d}} \cdot V_i}{\text{\tt exp}^{LSE}},
\]
where log-sum-exp $ LSE $ is defined as:
\[
    LSE = \log\sum_{i=0}^{R-1}\text{\tt exp}^{Q \cdot K_i^T / \sqrt{d}}.
\]
In Section \ref{sec:ring-pass-kv-algo}, we calculate the attention output and $LSE$ on each $CP$ rank $ k $:
\[
LSE_k^s, O_k^s = \text{\tt Attn}(Q_k, KV^s),
\]
with $ s = 0, 1, ..., N-1 $ on CP rank $ k $.

Similar to blocked {\tt softmax} computation in Flash Attention ~\cite{fa_v1,fa_v2} and the derivation process in \cite{juravsky2024hydragen}, we can get
\begin{equation}
\large
    O_k = \frac{\sum_{s=0}^{N-1}(O_k^s \times \text{\tt exp}^{LSE_k^s - LSE_k^{max}})}{\sum_{s=0}^{N-1} \text{\tt exp}^{LSE_k^s - LSE_k^{max}}},
    \label{eqn:merge_attn}
\end{equation}
where
$ LSE_{k}^{max} = \max_{s=0}^{N-1} LSE_k^s $.

In this way\footnote{Merge attention implementation is open sourced at \url{{https://facebookresearch.github.io/xformers/components/ops.html\#module-xformers.ops.fmha}}.}, we can combine attention output computed on different chunks of K/V for the same query to get attention on the whole K/V.


%





\section{Analytical Model Selection Considering All2All}
\label{sec:all2all}

{\tt pass-Q} merge attention requires an \textit{All2All} (Section \ref{sec:ring-pass-q-algo}), whereas in {\tt pass-KV} merge attention only needs to merge the partial attn results on local node (Section \ref{sec:ring-pass-kv-algo}). When {\tt pass-KV} communication is exposed, we want to compare the total of exposed {\tt pass-KV}'s communication time to the {\tt pass-Q}'s \textit{all2all}, which is the time to send partial attention output and partial attention softmax log-sum-exp (LSE) (Appendix \ref{sec:merge-attn}):




$$
Latency(\textit{All2All}) = (N-1) \cdot \frac {(D+1) \cdot T \cdot e}{BW}
$$

This means {\tt pass-Q} has better prefill latency only if:
$$
(N-1) \cdot \left(\frac{2 (T + P) D \cdot e \cdot \frac{N_{KV}}{N_H}}{BW} - \frac{4\cdot T \cdot D \cdot (T + P)}{N \cdot C}\right)
$$
$$ \ge (N-1) \cdot \frac {(D+1) \cdot T \cdot e}{BW}.
$$
Assuming $D \approx D+1 $, through algebraic rearrangement, we get:
\begin{equation}
\label{eqn:pass-kv-le-pass-q-a2a-1}
2 \cdot \frac{N_{KV}}{N_H} - \frac{4T \cdot BW}{N \cdot C \cdot e} \ge \frac{T}{T+P}
\end{equation}
Compared to \eqref{eqn:pass-q-vs-kv}, this shows that considering \textit{All2All} decreases the KV cache miss rate threshold for selecting \passq{}.

%


Algorithm \ref{alg:pass-kv-vs-pass-q-partial-adjusted-all2all} is the adjusted heuristic algorithm to select between \passkv{} and \passq{}, considering \textit{All2All} used in merge attention in \passq{}.

\begin{algorithm}[h]
    \caption{Pass-KV vs. Pass-Q Partial Prefill Heuristics}
   \label{alg:pass-kv-vs-pass-q-partial-adjusted-all2all}
\begin{algorithmic}
   \IF{
        $T \geq N \frac{C \cdot {N_{KV}} \cdot e}{2 \cdot {N_H} \cdot BW} $ or $ {\frac{T}{T+P}} \ge 2 \cdot \frac{N_{KV}}{N_H} - \frac{4T \cdot BW}{N \cdot C \cdot e} $
   }
   \STATE pass-KV
   \ELSE
   \STATE pass-Q
   \ENDIF
\end{algorithmic}
\end{algorithm}

\section{Heuristic based on empirical data}
\label{sec:heuristic}

\begin{figure}[t]
\centering
\includegraphics[width=8cm]{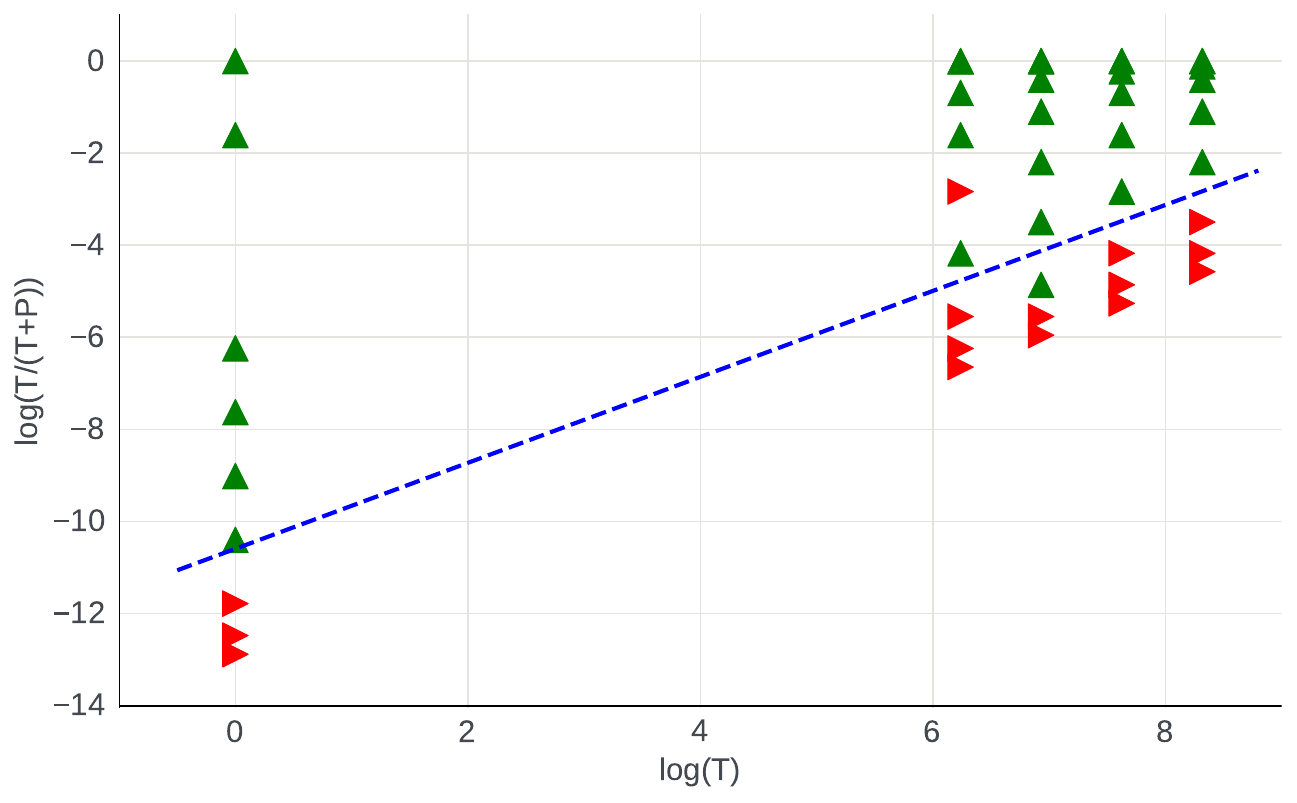}
\caption{A heuristic model using empirical data points. Green: prefer \passkv, Red: prefer \passq}
\label{fig:empirical-heuristic}
\end{figure}

For practical uses, we further establish a simplified heuristic to choose between {\tt pass-KV} and {\tt pass-Q} based on emprical data points. Particularly we collected data points for various combinations of $T$ and $T/(T+P)$, and establish an empirical formula:
$$
h(T, P) = \alpha \cdot \log(T) + \beta \cdot \log\left({T \over T+P}\right) + \gamma
$$
We prefer \passkv when $h$ evaluates to a positive value and prefer \passq otherwise. We fit empirical data points to this formula with parameters: $\alpha=-1.059$, $\beta=1.145$ and $\gamma=12.112$, as show in Figure~\ref{fig:empirical-heuristic}. One way to interpret the heuristic is that, for each particular $T$, there is a threshold for $T/(T+P)$ based on which we should switch from \passq to \passkv for best performances, and the threshold increases as $T$ increases.

Note that we do not expect the linear model to perfectly capture all cases, so some misclassifications are present due to variances and other factors, but the general trend is obvious. We inspected the misclassified data points, and they turned out to be the ones where the differences between the two strategies were relatively small ($<1\%$). In practice we can run this heuristic at the beginning of each round and get the best of both worlds.



\end{document}